On the analogy between the restricted primitive model and capacitor circuits. Part II: A generalized Gibbs-Duhem consistent extension of the Pitzer-Debye-Hückel term with corrections for low and variable relative permittivity

*Andrés González de Castilla,* Simon Müller, Irina Smirnova*

**corresponding author (**andres.gdc@tuhh.de**)**

*Hamburg University of Technology, TUHH, Institute of Thermal Separation Processes, Eißendorfer Straße 38 (O), 21073, Hamburg, Germany*

# Abstract

We present a novel, thermodynamically consistent modification of the Pitzer-Debye-Hückel term and its extension for concentration dependent density, molar mass and relative permittivity. This extension is validated for ionic liquids by comparison with a reference model from the literature and, in contrast to similar extensions, also applied to conventional salts with small spherical ions in aqueous, mixed and non-aqueous solvents. The central novelty is the inclusion of a modified parameter of closest approach, which improves the overall qualitative performance of the Pitzer-Debye-Hückel term over the complete relative permittivity range. Gibbs-Duhem consistency is retained in the modified extension and sample calculations for aqueous [BMIM][BF$_4$] and aqueous NaCl are provided. The novel, modified and extended term with concentration dependent properties is combined with the predictive COSMO-RS-ES model for the calculation of phase equilibria and activity coefficients in electrolytes with conventional salts. The performance of the COSMO-RS-ES model for predictions of salt solubility in fully non-aqueous media improves significantly upon introduction of concentration dependent properties within the long-range electrostatics. Modelling performance with the modified extended Pitzer-Debye-Hückel term outperforms modelling with the unmodified extension as well as with the conventional term with no extension. The correlated relative permittivity of the mixture is overestimated with respect to experimental values and kinetic depolarization effects provide a plausible explanation for this observation. Overall, our results support the consistent introduction of concentration dependent properties within the electrostatic theory in order to improve the modelling of electrolytes with particular emphasis on non-aqueous electrolytes.

# 1. Introduction

Electrolytes are ubiquitous in chemical processes and nature. Their thermodynamic modelling remains a challenging topic that has gained additional focus due to growing interest in salt-based deep eutectic mixtures, ionic liquids (ILs) and non-aqueous electrolytes in the chemical industry.[1–4] Modelling electrolyte systems in chemical engineering applications is a widespread practice that has found a strong application in excess Gibbs free energy ($g^E$) models[5–14] as well as equations of state (EoS).[15–20] All of which require a complement for long-ranged coulombic interactions between charged species, commonly given by Debye-Hückel theory,[21] the Mean Spherical Approximation (MSA)[22,23] or the Pitzer-Debye-Hückel term[24,25] and in many cases benefit from the use of a Born term.[17,26] The Pitzer-Debye-Hückel term and its extensions are most commonly applied to $g^E$-models like the electrolyte NRTL,[27–29] UNIFAC/UNIQUAC based electrolyte models[5,6,30–33] and COSMO-RS based electrolyte models.[8–12,34–37]

The thermodynamically consistent introduction of the dielectric constant in these electrolyte theories regained particular focus in the recent literature. The topic has been addressed by Shilov and Lyashchenko in their extended Debye-Hückel theory,[38–40] by Bülow et al.[16] for the ePC-SAFT EoS combined with an extended Debye-Hückel theory,[16,18] by Chang and Lin who combine the COSMO-SAC model with an extended Pitzer-Debye-Hückel term (E-PDH, also found in the literature[33] as ext-PDH) that was tailored for ILs in organic solvents[8,9] and tested by Ganguly et al.[33] with a UNIQUAC electrolyte model. The topic is also addressed by Walker et al.[41] with the SAFT-VR Mie EoS and by Lei et al.[42] from the Debye-Hückel perspective, among others.[15,20,43] The topic extends to Monte-Carlo based approximations as shown in the work Valiskó and Boda[44,45] or Chen and Panagiotopoulos,[46] where the relevance of transfer energetics from the Born solvation term play a central role. Thus, there seems to be a strong case for introducing a concentration dependent relative permittivity in the modelling of electrolyte systems.[47]

In the recent literature, the Pitzer-Debye-Hückel (hereafter, PDH) term has been extended and tailored for ionic liquid systems.[8] In the present work, we develop a modified PDH extension which includes a modified parameter of closest approach (denoted by $b''$) derived in our previous work.[48] It is the purpose of the present work to show that the inclusion of $b''$ retains Gibbs-Duhem consistency and can be extended for concentration dependent properties. For this purpose, PDH extensions are obtained through the differentiate down method[49–51] and tested with the Gibbs-Duhem relation. Qualitative equivalency to a conventional PDH extension for ionic liquids is discussed and, in contrast to other works,[8,9,33] our extension is also systematically applied to systems with salts for which the fused salt state is not available under normal conditions. Systematic application of our modified extended PDH (ME-PDH) term is done in combination with the predictive COSMO-RS-ES model for the calculation of mean ionic activity coefficients and phase equilibria in a wide variety of single salt systems with pure and mixed solvents. Our findings support the consistent introduction of concentration dependent properties in electrostatic models to improve the modelling of highly concentrated, non-aqueous electrolytes. Present and previous findings[48] also support the use of the modified parameter of closest approach $b''$ in both the extended and unextended PDH terms.

## 2. Theoretical Background

**2.1. Definition of over- and underscreening in electrolyte solutions**

As done in our previous work,[48] we refer to the negative deviation of the screening length from its theoretical Debye length value as *overscreening*. This phenomenon is explained by an increase of the leading screening parameter $\kappa$ with respect to the Debye screening parameter $\kappa_D$ that arises from specific ion-ion correlations (e.g. volume exclusion effects) and results in an increased effective charge of the ions.[52,53]

We refer to the anomalous positive deviation of the screening length from its theoretical value as *underscreening*. This phenomenon is given by an anomalous decrease of the leading screening parameter $\kappa$ with respect to its theoretical value $\kappa_D$, that arises from specific ion-ion correlations (e.g. dominant cation-anion coupling at high valences or very low dielectric constants) and results in a decreased effective charge of the ions.[54–57] In line with our previous publication,[48] the type of underscreening addressed relates to the invalidity of linearized electrolyte theories in a decreasing or very low relative permittivity.

**2.2. Basic Equations**

To the best of our knowledge, Lesser Blum was the first to suggest that linearized electrolyte theories can be expressed as a collection of spherical capacitors embedded in a dielectric medium.[58] We have recently demonstrated[48] how this analogy applies to various linearized electrolyte theories, including the novel Multiple Decay-Length Extension of the Debye-Hückel (MDE-DH) theory by Kjellander.[56] Linearized electrolyte theories can be represented as circuits of capacitors in series and parallel according to the following generalized form (shown here for the calculation of activity coefficients):[48,58]

$$\frac{\mu_i^{el}}{RT} = \ln(\gamma_i^{el}) = -\frac{z_i^2 e^2}{8\pi\varepsilon_0\varepsilon_r} \cdot \left(\frac{1}{C_i^X}\right) \qquad (1)$$

where $\mu_i^{el}$ stands for the electrostatic contribution to the chemical potential of species $i$, $C_i^X$ stands for an effective capacitance given by electrolyte theory $X$, $z_i e$ is the charge of particle $i$ and $\varepsilon_r$ stands for the relative permittivity of some arbitrary dielectric medium. It has been shown in the literature,[48,59] that the qualitative behavior of the logarithmic activity coefficient $\ln(\gamma_i^{el})$ from different theories performs similarly when linearly scaling of the distance of closest approach (hard-core collision diameter) $a = r_+ + r_-$, where $r_+$ and $r_-$ are ion specific as discussed by Ribeiro et al.[60] This similarity of performance was thoroughly evaluated for the first time by Maribo-Mogensen et al.[59] In their work, Debye-Hückel theory yielded very similar qualitative results when compared to the unrestricted MSA if the hard-core collision diameter was scaled by 5/6 of the value applied in MSA. The same effect was observed for the chemical potential in our previous work when the hard-core collision diameter in the PDH term is equal to 3/2 of the hard-core collision diameter applied in MDE-DH theory.[48]

The scaling of the hard-core collision diameter is a qualitative compensation for the absence (or incomplete description) of volume exclusion effects (a type of overscreening) in some linearized formulations of the primitive model.[48,59] In our previous work,[48] it was suggested that if a scaling of the diameter can qualitatively approximate volume exclusion effects in DH theory, then it can also do so for underscreening. It was then proposed that this is what the recommended literature values for the parameter of closest approach of the Pitzer equations actually stand for. Based on these ideas, a theory with some hypothetical screening length $\kappa^H$ was assumed to be able to describe underscreening and that it could be approximated in Debye-Hückel theory when scaling the diameter by some function $f(\kappa^H, \kappa_D, a)$, as shown in Figure 1.

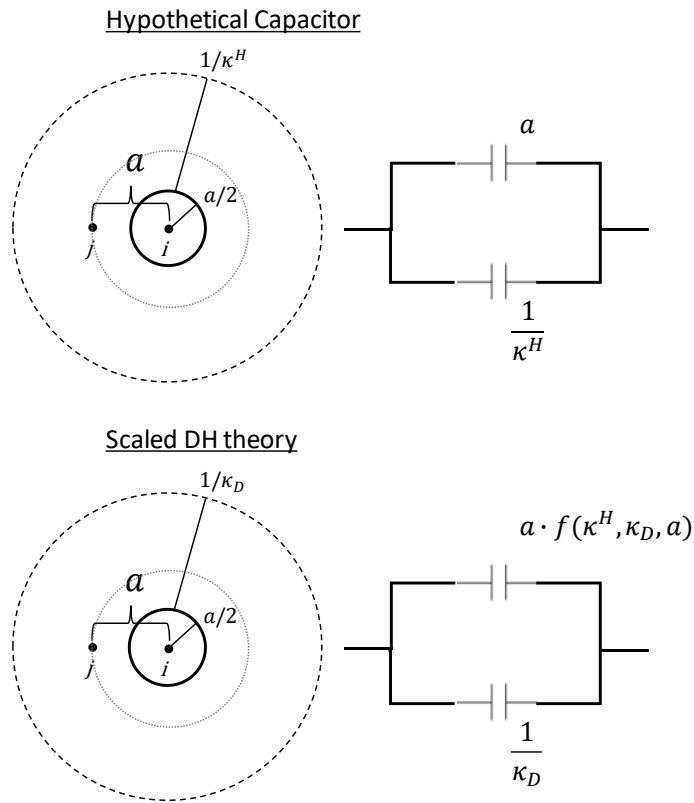

*Figure 1. Hypothetical capacitor for underscreening and Debye-Hückel theory with a scaled diameter.*

Based on a semi-linear extrapolation of a lower order approximation[55,56] for the Dressed Ion Theory,[61,62] the following expression was taken for $\kappa^H$:

$$\frac{\kappa_D}{\kappa^H} = \sqrt{1 - \alpha_1 \Lambda^2} \qquad (2)$$

where $\Lambda$ is a coupling factor defined as the product of the Debye length $\kappa_D$ and the Bjerrum length $\lambda_B$ and $\alpha_1$ is a constant given by:[55]

$$\alpha_1 = -\frac{K_{assoc}(T, \varepsilon_r, a)}{8\pi z_i^2 N_A \lambda_B^3} \qquad (3)$$

Based on these assumptions and equations, the following modification for the closest approach parameter was introduced for the mole fraction basis:[48]

$$b'' = a\sqrt{\frac{9 d_s N_A e^2}{2 M_s \varepsilon_0 \varepsilon_s k_B T}} + \sqrt{\frac{1}{I_x} + \frac{d_s}{M_s} \cdot 8\pi N_A (\alpha_{sc} a)^3 \sum_{n=1}^{\infty} \frac{B_{sc}^{2n+2}}{(2n+2)! \cdot (2n-1)}} - \frac{1}{\sqrt{I_x}} \qquad (4)$$

where subscript $s$ denotes solvent properties for density $d_s$, molar mass $M_s$ and relative permittivity $\varepsilon_s$, $I_x$ stands for ionic strength in the mole fraction scale, the scaling factor $\alpha_{sc}$ takes a value close to 30 to approximate literature values of the parameter of closest approach,[48,63] and $B_{sc}$ is a scaled fluid specification calculated as follows:

$$B_{sc} = \frac{\lambda_B}{(\alpha_{sc} a)} = \left(\frac{e^2}{4\pi \varepsilon_0 \varepsilon_s k_B T}\right)\left(\frac{1}{\alpha_{sc} a}\right) \qquad (5)$$

where $\lambda_B$ is the Bjerrum length. The use of $b''$ in the PDH term is to be tested and adapted for consistency.

In Pitzer's work, the original parameter of closest approach (denoted by $b'$) is an input for the excess Gibbs free energy of long-ranged electrostatic origin:[25]

$$\frac{G^{DH}}{RT} = -\frac{4 A_x n_T I_x}{b'} \ln(1 + b'\sqrt{I_x}) \qquad (6)$$

$$A_x = \frac{1}{3}\left(\frac{2\pi N_A d_s}{M_s}\right)^{0.5}\left(\frac{e^2}{4\pi \varepsilon_0 \varepsilon_s k_B T}\right)^{1.5} \qquad (7)$$

were the total mole number is $n_T = \sum_i n_i$ and $A_x$ is the Debye-Hückel parameter in mole fraction scale. By partial differentiation and applying $b'$, Pitzer obtained his widely known contribution for the activity coefficient of charged and uncharged species expressed by equation (8) for the unsymmetric reference state:[64]

$$\ln(\gamma_i^{PDH}) = \frac{\partial}{\partial n_i}\left\{\frac{G^{DH}}{RT}\right\} = -A_x\left[\frac{2z_i^2}{b'}\ln(1+b'\sqrt{I_x}) + \frac{z_i^2\sqrt{I_x} - 2I_x^{3/2}}{1+b'\sqrt{I_x}}\right] \quad (8)$$

where $n_i$ stands for moles of any species $i$. If salt-free properties are applied for density, relative permittivity and molar mass, then the parameter of closest approach is a constant (e.g. a recommended value[25] of $b' = 14.9$ for water). Further derivation leads to equation (9):

$$\frac{\partial \ln(\gamma_i^{PDH})}{\partial n_i} = -A_x\left[\frac{3z_i^2 I_x^{-1/2} + 2b'z_i^2 - 6\sqrt{I_x} - 4b'I_x}{2(1+b'\sqrt{I_x})^2}\right]\left(\frac{\partial I_x}{\partial n_i}\right) \quad (9)$$

with the definition $\frac{\partial I_x}{\partial n_i} = \frac{z_i^2}{2n_T} - \frac{I_x}{n_T}$ (where again subscript $i$ stands for any species in the mixture).

For a single solvent system, one can test Gibbs-Duhem consistency with equation (9) and the following reformulation of the Gibbs-Duhem relation with sufficiently small concentration steps:

$$\sum_j x_j \frac{\Delta \ln(\gamma_j^{PDH})}{\Delta x_i} = -x_s \frac{\Delta \ln(\gamma_s^{PDH})}{\Delta x_i} \quad (10)$$

where the subscript $j$ stands for ionic components and subscript $s$ stands for solvent (neutral species) and subscript $i$ stands for any arbitrary species. One can thereby show that equation (8) is Gibbs-Duhem consistent, which is an expected result based on the work of Pitzer.[24,50,65] Retaining such consistency should be a central goal to achieve thermodynamic coherency in any modification or extension of the PDH term. It is the purpose of the present work to provide Gibbs-Duhem consistent reformulations of the PDH term using $b''$ instead of $b'$, provide the corresponding extensions for concentration dependent density, molar mass and relative permittivity and to systematically test these expressions.

## 3. Development of a modified extended PDH term
### 3.1. Adaptation of the modified parameter of closest approach

In order to extend the PDH term for mean ionic activity coefficients (MIAC) within the framework of our recently published[48] modification for the closest approach parameter, shown in equation (4), the thermodynamic consistency of this modification needs to be tested when applied to equation (8).

One may apply equation (4) directly in equation (8) as done in our previous work.[48] This provides a systematic correction when assuming full dissociation with very low $\varepsilon_s$ values and provides activity coefficients that approximate the MDE-DH theory when applying the same hard-core collision diameter as in MDE-DH at higher $\varepsilon_s$ values (approx. $\varepsilon_s > 30$). Nevertheless, one can show that the use of equation (4) as parameter of closest approach in equation (8) incurs in a small Gibbs-Duhem consistency as salt concentration rises (corresponding test shown in the Supporting Information). An analysis of the derivatives can demonstrate that this is due to the terms that contain $I_x$ in equation (4). These terms cancel each other out at infinite dilution, have a negligible contribution to the value of $b''$ and incur into a small inconsistency. Based on these grounds, $I_x$ can be eliminated without affecting the performance of $b''$ to approximate literature values (refer to the Supporting Information). Thereby one can rewrite a considerably simplified reformulation of equation (4) in terms of $b'$ as follows:

$$b'' = b_{(0)} + b_{(1)} \tag{11}$$

$$b_{(0)} = \omega_{(0)} b' \tag{12}$$

$$b_{(1)} = \left(\omega_{(1)} \frac{\lambda_B}{a}\right)^{3/2} b' \tag{13}$$

$$b' = a \sqrt{\frac{2 d_s N_A e^2}{M_s \varepsilon_0 \varepsilon_s k_B T}} \tag{14}$$

where $\omega_{(0)}$ and $\omega_{(1)}$ are universal, fixed parameters with recommended values of $3/2$ and $1/9$, respectively.[48,63] This reformulation makes the modified parameter of closest approach easily differentiable and retains Gibbs-Duhem consistency when applied to equation (8) using salt free properties $M_s, d_s$ and $\varepsilon_s$. Thus, it is directly applicable to the Pitzer equations for the calculation of $\gamma_i$ and osmotic coefficients (molality equivalents and sample calculations included in the Supporting Information). The modification provides a hard-core collision diameter dominated term $b_{(0)}$ which retains the same interpretation from Pitzer:[24,65] "a parameter related to the diameter of the ions"; and a Bjerrum length dominated term $b_{(1)}$ that could represent the long-ranged electric fields of opposite charge partially cancelling each other out as the relative permittivity attains very low values. A different interpretation is that the hard-core collision diameter is scaled by an empirical factor $\left(\omega_{(0)} + \left(\omega_{(1)} \frac{\lambda_B}{a}\right)^{3/2}\right)$ that strongly depends on the relative permittivity.

### 3.2. Generalized extension of the PDH and modified PDH terms

We now proceed to introduce concentration dependent molar mass $M_m$, density $d_m$ and relative permittivity $\varepsilon_m$ (where subscript $m$ stands for mixture), which replace the salt-free properties (subscript $s$) in the calculation of $A_x, b'$ and $b''$. Details of the derivation of equation (6) with concentration dependent properties to obtain PDH extensions is included in the Supporting Information.

A fully extended PDH term (E-PDH) with the traditional parameter of closest approach $b'$ from equation (14) develops into the following expression that is consistent with the derivation from Chang and Lin:[8]

$$\ln(\gamma_i^{E-PDH}) = -A_x \Bigg\{ \frac{2z_i^2}{b'} \ln(1 + b'\sqrt{I_x}) + \frac{z_i^2 I_x^{1/2} - 2I_x^{3/2}\left(\frac{M_i}{M_m}\right)}{(1 + b'\sqrt{I_x})}$$

$$- \frac{2I_x^{3/2} n_T}{(1 + b'\sqrt{I_x})} \left[\frac{1}{\varepsilon_m}\frac{\partial \varepsilon_m}{\partial n_i} - \frac{1}{d_m}\frac{\partial d_m}{\partial n_i}\right] \qquad (15)$$

$$- \frac{4I_x \ln\left(1 + b'I_x^{1/2}\right) n_T}{b'} \frac{1}{\varepsilon_m}\frac{\partial \varepsilon_m}{\partial n_i} \Bigg\}$$

and a fully modified extended PDH term (ME-PDH) with our modified parameter of closest approach $b''$ from equation (11) develops correspondingly into the following:

$$\ln(\gamma_i^{ME-PDH}) = -A_x \Bigg\{ \frac{2z_i^2}{b''} \ln(1 + b''\sqrt{I_x}) + \frac{z_i^2 I_x^{1/2} - 2I_x^{3/2}\left(\frac{M_i}{M_m}\right)}{(1 + b''\sqrt{I_x})}$$

$$- \frac{2I_x^{3/2} n_T}{(1 + b''\sqrt{I_x})} \left[\left(1 + \frac{3b_{(1)}}{b''}\right)\frac{1}{\varepsilon_m}\frac{\partial \varepsilon_m}{\partial n_i} - \frac{1}{d_m}\frac{\partial d_m}{\partial n_i}\right] \qquad (16)$$

$$- \frac{2I_x \ln\left(1 + b''I_x^{1/2}\right) n_T}{b''} \left(2 - \frac{3b_{(1)}}{b''}\right)\frac{1}{\varepsilon_m}\frac{\partial \varepsilon_m}{\partial n_i} \Bigg\}$$

In both E-PDH and ME-PDH, the calculation of $A_x$ and $b'$ or $b''$ are now performed with mixture properties ($M_m$, $d_m$, $\varepsilon_m$) instead of salt-free properties. One may note that setting $\omega_{(0)} = 1$ and $\omega_{(1)} = 0$ reduces the ME-PDH term of equation (16) to the E-PDH term of equation (15).

If point charges embedded in a dielectric medium of constant volume are taken, which corresponds to the strict Poisson-Boltzmann definition, concentration dependent properties may be introduced. For instance, in the linear regime of the concentration dependent relative permittivity (dielectric decrement) the property $\varepsilon_m$ can have the form $\varepsilon_m = -A * I_x + \varepsilon_s$ and $\frac{\partial \varepsilon_m}{\partial n_i} = -A \frac{\partial I_x}{\partial n_i}$ with $-A$ as an empirical slope and $\varepsilon_s$ as the salt free permittivity of the solvent(s) in question. The same applies to the density in a linear regime: $d_m = C * I_x + d_s$ and $\frac{\partial d_m}{\partial n_i} = C \frac{\partial I_x}{\partial n_i}$. However, non-linear functions for $\varepsilon_m$ and $d_m$ based on experimental data, phenomenological models, theoretical models or mixing rules are also applicable as long as the partial derivatives of $\varepsilon_m$ and $d_m$ with respect to each species are accessible or accurately approximated.

### 3.3. Exemplarily development of equation (16) with volume fraction mixing rules

Some simple alternatives for the specific case of ionic liquids have been tested in the literature. For instance, mass and mole fraction mixing rules were applied for the first time with the E-PDH term by Chang and Lin[8] and their expressions have already been successfully applied in the literature.[8,9,14,51] Based on the work of Koeberg et al.[66], Liu et al.[67] and recent modelling strategies with the ePC-SAFT,[16,68] it also seems reasonable to apply volume fraction mixing rules (fully dissociated salt basis) as follows:

$$\varepsilon_m = \sum_i \phi_i^V \varepsilon_i \qquad (17)$$

$$d_m = \sum_i \phi_i^V d_i \qquad (18)$$

where $\phi_i^V$ stands for volume fraction of component $i$. In order for the volume fraction mixing rules on a fully associated IL basis (solvent + IL, 2 species) from Liu et al.[67] and Koeberg et al.[66], and on a fully dissociated IL basis (solvent + dissociated IL, 3 species) to be equivalent, the simplifications $\varepsilon_{(+)} = \varepsilon_{(-)} = \varepsilon_{IL}$ and $d_{(+)} = d_{(-)} = d_{IL}$ are required and the molar mass of the salt must be distributed between the ions as $M_{(+)} = M_{(-)} = \frac{M_{IL}}{v}$. These are reasonable choices given the questionable validity of single ion properties.[69] Doing so is also consistent with the DH framework, which assumes that all ions and counter-ions in the ion cloud are exactly equal except for opposing charges. By applying volume fraction mixing rules with the aforementioned assumptions, equation (16) develops as follows:

$$\ln(\gamma_i^{ME-PDH})_\phi = -A_x \left\{ \frac{2z_i^2}{b''} \ln(1+b''\sqrt{I_x}) + \frac{z_i^2 I_x^{1/2} - 2I_x^{3/2}\left(\frac{M_i}{M_m}\right)}{(1+b''\sqrt{I_x})} \right.$$

$$- \frac{2I_x^{3/2}}{(1+b''\sqrt{I_x})}\left[\left(\frac{V_i}{V_m}\right)\left(1-\frac{d_i}{d_m}\right) + \left(\frac{V_i}{V_m}\right)\left(1+\frac{3b_{(1)}}{b''}\right)\left(\frac{\varepsilon_i}{\varepsilon_r}-1\right)\right] \qquad (19)$$

$$\left. - \frac{2I_x \ln\left(1+b''I_x^{1/2}\right)}{b''}\left(2-\frac{3b_{(1)}}{b''}\right)\left(\frac{V_i}{V_m}\right)\left(\frac{\varepsilon_i}{\varepsilon_r}-1\right) \right\}$$

where subscript $\phi$ denotes the use of volume fraction and $V_i$ stands for molar volume of species $i$ and $V_m$ is the total molar volume of the mixture. It is noteworthy to mention that by applying a mass fraction mixing rule for $\varepsilon_m$, a mole fraction mixing rule for density ($d_m^{-1} = \sum_i x_i d_i^{-1}$) and setting $\omega_{(1)} = 0$ and $\omega_{(0)} = 2.5$, then the equation reduces to the result from Chang and Lin,[8] which is the first full extension that stemmed from E-PDH. A minor difference is that Chang and Lin[8] assume $M_{ion} = M_{salt}$ for all ions (in contrast to $M_{(+)} = M_{(-)} = \frac{M_{IL}}{v}$). A central difference is applying $b''$ instead of the scaling factor of 2.5 which is referred to as an "ion pair formation parameter" used by Chang and Lin[8] in $b'$.

Finally, volume fraction mixing rules are directly interchangeable with mass fraction mixing rules by replacing $\phi_i^V$ with $w_i = \sum_i \frac{x_i M_i}{\sum_i x_i M_i}$ and replacing $\frac{V_i}{V_m}$ with $\frac{M_i}{M_m}$ in both E-PDH or ME-PDH.

## 4. Methods

The calculations performed in this work are divided into three sections. The first part compares diverse calculations with several versions of the PDH term for an ionic liquid system in order to demonstrate qualitative equivalency with Chang and Lin´s extension[8] from the literature and to demonstrate where the correction applied by $b_{(1)}$ becomes relevant. The second part deals with a purely theoretical calculation for aqueous NaCl using the modified extended PDH term, a Born term and an unrestricted hard-sphere term to demonstrate theoretical applicability in small spherical ions. The third part compares the modelling performance of the PDH extensions in combination with the COSMO-RS-ES model[10–12] showing the applicability for the calculation and prediction of activity coefficients and phase equilibria in diverse single salt systems in aqueous, mixed and non-aqueous solvents over a wide concentration range.

**4.1. Long-Range term comparison between PDH extensions for an IL: Methods**

For a qualitative comparison with an aqueous ILs, equation (19), which is the equivalent of equation (16) applying volume fraction mixing rules, is applied using the properties shown in Table 1.

*Table 1. Pure component properties applied for the calculations of aqueous [BMIM][BF$_4$].*

| Property | Water | [BMIM]$^+$ | [BF$_4$]$^-$ |
|---|---|---|---|
| $d_i\ [kg/m^3]$ | 998 | 1201.5[a] | 1201.5[a] |
| $\varepsilon_i$ | 78.34 | 13.9[a] | 13.9[a] |
| $M_i\ [kg/mol]$ | 0.018015 | 0.13922[b] | 0.08681[b] |

[a] Ionic liquid experimental values taken from data collected by Marcus[69] [b] Applied as (0.13922 + 0.08681)/2 for each ion due to equivalence with volume fraction mixing rules on the fully associated basis. The distance of closest approach (hard-core collision diameter) was set to $a = 4$ Å.

The results are compared with the model from Chang and Lin[8] in order to show qualitative equivalency in the ionic liquid range between both PDH extensions as well as equation (15) applying volume fraction mixing rules (the extension of the original PDH). Subsequently the effect of the term $b_{(1)}$ is discussed and Gibbs-Duhem consistency is evaluated for all cases by means of equation (10). These calculations, including those with the extension from Chang an Lin,[8] are done at full dissociation and treat the molar mass of each ion as $\frac{M_{IL}}{v}$ in order to keep the same basis. An additional evaluation treating a fraction of the ionic liquid as neutrally charged molecular pseudo-solvent is also included and briefly discussed.

### 4.2. Theoretical Primitive Model comparison for aqueous NaCl: Methods

As an initial test for spherical ions, aqueous NaCl at room temperature is applied with $a = 2.79$ Å based on the ionic radii reported by Marcus[70] with $M_{(+)} = 22.99 \, mol \, g^{-1}$ and $M_{(-)} = 35.45 \, mol \, g^{-1}$. The density and relative permittivity of the mixture are estimated as functions of $I_x$ by equations (20) and (21) which are based on the work of Gates and Wood[71] and the work of Buchner et al.[72] respectively:

$$d_m \, [kg/m^3] = 2225.7 I_x + 998 \tag{20}$$

$$\varepsilon_m = 3620.1 I_x^2 - 740.24 I_x + 78.35 \tag{21}$$

Equations (20) and (21) are valid up to a concentration of 3 molal.

The calculations are performed with the extension of the original PDH term with $b'$ from equation (15) and the extension of the PDH term with the modified parameter of closest approach $b''$ from equation (16) from the present work as electrostatic terms. Experimental data from Hamer and Wu[73] are also presented for qualitative comparison. Similar to our previous work,[48] the electrostatic terms are combined with the Born term (BT) applying the same Born radii from Valiskó and Boda[44] ($R_{Na^+}^{Born} = 1.62$ Å and $R_{Cl^-}^{Born} = 2.26$ Å) for mathematical equivalency with their Monte-Carlo based model and with an unrestricted hard-sphere (UHS) contribution taken from Ebeling and Scherwinski as follows:[74]

$$\ln(\gamma_\pm) = \ln(\gamma_\pm^{PDH}) + \ln(\gamma_\pm^{Born}) + \ln(\gamma_\pm^{UHS}) \tag{22}$$

The pertaining equations for $\ln(\gamma_\pm^{Born})$ and $\ln(\gamma_\pm^{UHS})$ are included in the Supporting Information and also described in our previous publication.[48]

### 4.3. Universal application for common salts with the COSMO-RS-ES model: Computational Details

The COSMO-RS-ES model[10–12] is a predictive electrolyte model which combines tailored universal ion contributions based on the COSMO-RS theory[75,76] with a PDH term for electrostatics in the usual form:

$$\ln(\gamma_i) = \ln(\gamma_i^{SR,COSMO-RS}) + \ln(\gamma_i^{LR,PDH}) \tag{23}$$

As PDH term, the three versions given by equations (8), (15) and (16) are tested to assess their impact.

Details of the COSMO-RS-ES model are found in previous publications.[10,12] The short-range COSMO-RS contributions come from ion – ion and ion – solvent specific energy interactions in terms of the surface charge of molecular segments ($\sigma$) and global parameters (A, B, C, D, E), as shown in Table 2:

*Table 2. COSMO-RS-ES tailored ion equations as described in Müller et al.[12]*

| Interaction | Misfit Factor | Ionic Interaction Energy Term |
|---|---|---|
| cation – $H_2O$ | $A_1$ | $E^{ion}_{cat-H_2O} = \frac{a_{eff}}{2} B_1 \sigma_{cat} \max(0, \sigma_{H_2O} - \sigma_{HB})$ |
| cation – org. mol. | $A_2$ | $E^{ion}_{cat-om} = \frac{a_{eff}}{2} B_2 \sigma_{cat} \max(0, \sigma_{om})$ |
| cation - halide | 0 | $E^{ion}_{cat-hal} = \frac{a_{eff}}{2} B_3 \min(0, \sigma_{cat}(1 - D_1|\sigma_{cat}|^{E_1})) \sigma_{hal}$ |
| cation – polyat. an. | 0 | $E^{ion}_{cat-pa} = \frac{a_{eff}}{2} B_4 \min(0, \sigma_{cat}(1 - D_1|\sigma_{cat}|^{E_1})) \max(0, \sigma_{pa})^{E_2}$ |
| halide - $H_2O$ | $A_3$ | $E^{ion}_{hal-H_2O} = \frac{a_{eff}}{2} B_5 \min(0, \sigma_{H_2O} + \sigma_{HB}) \max(0, \sigma_{hal})$ |
| halide – org. mol. | $A_4$ | $E^{ion}_{hal-om} = \frac{a_{eff}}{2} B_6 \min(0, \sigma_{om} + C_1) \max(0, \sigma_{hal} - C_2)^{E_3}$ |
| polyat. an. - $H_2O$ | $A_5$ | $E^{ion}_{pa-H_2O} = \frac{a_{eff}}{2} B_7 \min(0, \sigma_{H_2O} + \sigma_{HB}) \max(0, \sigma_{pa})$ |
| polyat. an. - org. mol. | $A_6$ | $E^{ion}_{pa-om} = \frac{a_{eff}}{2} B_8 \min(0, \sigma_{om} + C_1) \max(0, \sigma_{pa} - C_3)$ |

Value of $\sigma_{HB}$ remains fixed at 0.0085 [e/Å²]

The training set consists of aqueous mean ionic activity coefficients (MIACs) of salts in aqueous, aqueous mixed and non-aqueous solvents ($N_{DP}$ = 3865) up to concentrations of 15 molal and liquid-liquid equilibria (LLE) in aqueous mixed solvents ($N_{DP}$ = 1056). Experimental measurements of the relative permittivity of electrolytes in water, methanol and ethanol ($N_{DP}$ =259) are also included. The cations considered are alkali and ammonium ions with halides as monoatomic anions and $SO_3^{2-}$, $SO_4^{2-}$, $NO_3^-$, $ClO_4^-$, $H_2PO_4^-$, $HPO_4^{2-}$ and $S_2O_3^{2-}$ as polyatomic anions. Given that at present the COSMO-RS-ES model deals with conventional salts and not ionic liquids, the subscript "IL" is exchanged for "salt" for the hypothetical fused salt state.

Model parameters are optimized with a Levenberg-Marquardt algorithm with objective functions that have been described in previous works.[10,11] The selected measure for accuracy is the average absolute deviation (AAD) for the number of data points ($N_{DP}$) per system type:

$$AAD_{MIAC} = \frac{1}{N_{DP}} \sum_i \left| ln\left(\gamma^{(m),*,exp}_{\pm,i}\right) - ln\left(\gamma^{(m),*,calc}_{\pm,i}\right) \right| \qquad (24)$$

$$AAD_{LLE} = \frac{1}{N_{DP}} \sum_i \left| \ln(K_{\text{salt},i}^{OS,exp}) - \ln(K_{\text{salt},i}^{OS,calc}) \right| \qquad (25)$$

$$AAD_{Perm} = \frac{1}{N_{DP}} \sum_i \left| \varepsilon_r^{exp} - \varepsilon_r^{calc} \right| \qquad (26)$$

where $K_{\text{salt}}^{OS}$ is the partition coefficient of a salt between an salt-rich phase $S$ and an organic phase $O$ (in mole fraction convention); the subscript $(m)$ denotes the molality convention.

The objective function for the newly introduced aqueous electrolyte dielectric decrement data is given by:

$$OF_{Perm} = 0.4 \sum_i \left( \varepsilon_r^{exp} - \varepsilon_r^{calc} \right)^2 \qquad (27)$$

where 0.4 is a weighting factor. The weighting factors of the objective functions for aqueous and mixed MIACs of halide salts were set to 15 and for non-aqueous MIACs a value of 2. For salts with polyatomic ions, the weighting factors are 5 and 1 for MIACs in aqueous/aqueous-mixed mixed and non-aqueous solvents, respectively. The LLEs required no weighting factor.

Data that does not belong to the training set are considered predictions in this work. Predictions are performed for salt solubility data (SLE, 836 data points) which are calculated as follows:

$$\ln\left(\gamma_{\pm,i,other}^{+,expected}\right) = \ln\left(x_{\pm,ref} \cdot \gamma_{\pm,ref}^{+}\right) - \ln(x_{\pm,experiment}) \qquad (28)$$

where the subscript $ref$ stands for a reference solvent, which is taken as water for all cases and $\gamma_{\pm,i,other}^{+,expected}$ therefore stands for a transfer activity coefficient between the reference solvent and the modelled (other) solvent. Further details for this calculation can be found in Müller et al.[12]

To use the dielectric constant database, the raw experimental data must preprocessed in order to express it in terms of mole fraction. The general density model for aqueous electrolytes from Nguyen et al.[77] was applied for the pertaining change to mole fraction when the data was given in the molar concentration scale. The vast majority of this dielectric decrement data is for salts in water. For non-aqueous solvents there is no general density model available, but the assumption molarity = molality was considered valid given that the experimental data in these few cases is given for low salt concentrations.

The COSMO-RS-ES model[10–12] has been tested with a renormalized ionic strength in the PDH term but has not yet been evaluated with a consistent inclusion of mixture properties at high concentrations. This is performed in the present work for the first time. The present work applies the same database as in previous works[10–12] with additional datasets for MIAC values in mixed aqueous and non-aqueous solvents for 1:1 and 1:2 alkali and ammonium salts. Application to large polyatomic cations and ionic liquids with concentration dependent properties is object of future work.

The molar mass of the mixture has a clear definition. However, for the other properties in the present evaluation (density and relative permittivity) new generalized treatments are introduced. For the density of the mixture, a mass fraction mixing rule with the averaged properties of the salt-free and solvent-free states is applied:

$$d_m = \sum_j \frac{x_j \bar{M}_j \bar{d}_{salt}}{\sum_i x_i \bar{M}_i} + \sum_s \frac{x_s \bar{M}_s \bar{d}_s}{\sum_i x_i \bar{M}_i} \tag{29}$$

where subscripts $s$ salt-free medium and subscript $j$ stands for ions (solvent-free medium) and subscript $i$ stands for all species. The average molar mass of the solvent-free state is the same for all ions and given by the molar mass of the hypothetical molten salt and the stoichiometric coefficients: $\bar{M}_j = \frac{M_{salt}}{\nu}$. The average properties $\bar{M}_s$ and $\bar{d}_s$ are used for the salt-free state with $\bar{d}_s$ calculated with mass fraction on a salt-free basis for mixed solvents. The hypothetical density $\bar{d}_{salt}$ applied to all ions is given by:

$$\bar{d}_{salt} = 1076.48004 \cdot M_{salt}^{0.1553} \tag{30}$$

with $M_{salt}$ given in g/mol. Qualitative validation for aqueous systems is provided in the Supporting Information. The present work assumes this correlation is qualitatively acceptable for all systems.

For the relative permittivity of the mixture, we have combined the approaches from Inchekel et al.[78] and Bülow et al.[16] with some minor modifications as follows (details found in the Supporting Information):

$$\varepsilon_m = \left[\frac{\varepsilon_s^{\phi_{sf}}}{1 + \alpha_{salt} \sum_j x_j}\right]\left(\sum_s x_s\right) + \varepsilon_{salt}\left(\sum_j x_j\right) \tag{31}$$

where subscripts $s$ stands for solvent (or salt-free medium in the case of mixed solvents) and subscript $j$ stands for ions and $\varepsilon_s^{\phi_{sf}}$ stands for salt-free permittivity calculated on a volume fraction basis. The term $\varepsilon_{salt}$ is an effective extrapolation of the hypothetical fused salt state for which a constant value of $\varepsilon_{salt} = 13$ was initially selected as an educated guess based on the typical range for ionic liquids reported by Marcus.[69] It is assumed that the relative permittivity of conventional molten salts does not deviate significantly from this value for the purpose of the mixing rule. The parameter $\alpha_{salt}$ is a generalized pre-factor taken as a single universal parameter for now. Equation (31) therefore roughly captures a generalized behavior of $\varepsilon_m$ for a wide variety of systems with small ions with localized charges by means of a single parameter. From the mathematical form, one may observe that $\sum_j x_j \to 0$ for the salt-free medium with $\frac{\partial \varepsilon_m}{\partial n_s} \to 0$ and $\varepsilon_m \to \varepsilon_s^{\phi_{sf}} = \sum_s \varepsilon_s \phi_s^{sf}$, whereas for the hypothetical fused salt at $\sum_j x_j \to 1$ one obtains $\frac{\partial \varepsilon_m}{\partial n_j} \to 0$ and $\varepsilon_m \to \varepsilon_{salt}$ regardless of the value of $\varepsilon_s$.

The hard-core collision diameter in the COSMOR-RS-ES model is calculated as a scaled sum of the COSMO radii given by $a = f_{sc}\left(r_{(+)}^{COSMO} + r_{(-)}^{COSMO}\right)$ where $f_{sc}$ is a scaling factor left as an open parameter.[12] Finally, for the calculations with equation (16) as long-range term, the parameters $\omega_{(0)}$ and $\omega_{(1)}$ from equations (12) and (13) remain fixed with their recommended values of $3/2$ and $1/9$, respectively. It must be pointed out that both $f_{sc}$ and $r_{(+)}^{COSMO}$ are parameters for a general average of the hard-core collision diameter.

## 5. Calculations and Results
**5.1. Long-Range term comparison between PDH extensions for an IL: Results**

In Figure 2, results from equations (15) and (16) applying volume fraction mixing rules for the selected aqueous ionic liquid system are presented. Calculations with the extended version from Chang and Lin[8] are also shown. One detail must be pointed out: with the exception of molar mass of the ions given by $\frac{M_{IL}}{v}$ in all cases (as previously stated), the model from Chang and Lin[8] is used as published. In their publication the authors selected different mixing rules for the density and the relative permittivity of the mixture.

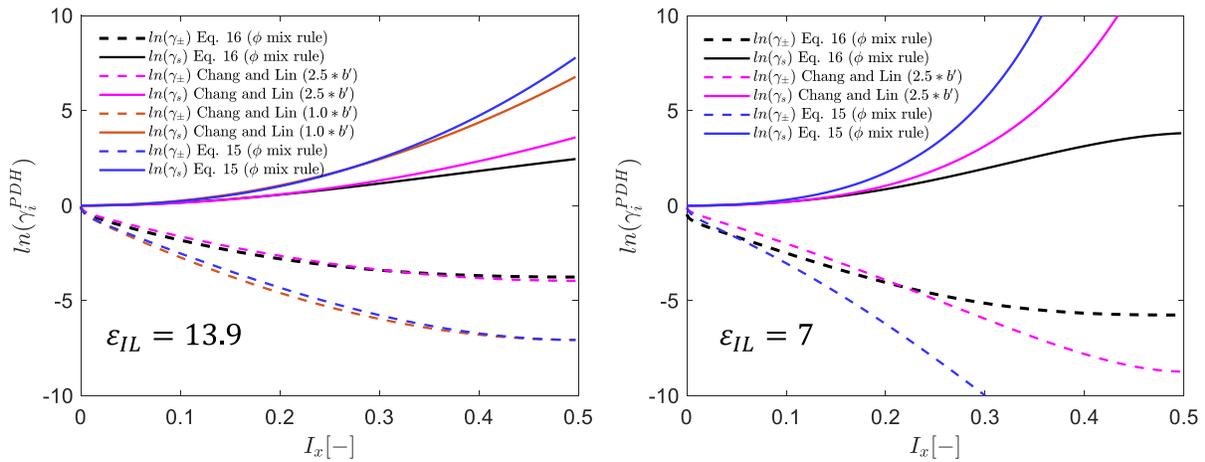

*Figure 2. Contributions from $\ln(\gamma_\pm^{PDH})$ and $\ln(\gamma_s^{PDH})$ for aqueous [BMIM][BF$_4$] at room temperature from infinite dilution to pure IL with concentration dependent molar mass, density and relative permittivity. For calculation purposes the hard-core collision diameter was set at $a = 4$ Å. Left: assumption of $\varepsilon_{IL} = 13.9$. Right: assumption of $\varepsilon_{IL} = 7$.*

It can be observed in Figure 2 (left) that there is clear qualitative agreement between equations (15) and (16), and the model from Chang an Lin[8] with $b = 2.5b'$ and $b = b'$, respectively. The calculation of the electrostatic contribution to $\ln(\gamma_\pm)$ is virtually equal and minor qualitative differences are observed in the calculation of the electrostatic contribution to the solvent: $\ln(\gamma_s)$. The latter qualitative differences are attributed to the use of different mixing rules to calculate the relative permittivity of the mixture $\varepsilon_r$. It must be pointed out that while the use of volume fraction mixing rules seems more physically meaningful,[16,66,79] molar mass and molar volume usually correlate. The natural consequence is that the differences in the resulting calculations seem negligible at first glance.

The modified extended PDH (equation (16)) and the original model from Chang an Lin[8] (with $2.5b'$) are in general qualitative agreement. This is an expected result: the value of 2.5 is an average based on recommended values from Barthel[80] and $b''$ (equation (11)) is a semi-empirical function that intends to reproduce these recommended values. However, the central advantage of $b''$ is that it will be lower than $2.5b'$ at higher relative permittivity values (refer to the discussion on Figure 5) and it will attain values higher than $2.5b'$ if the relative permittivity of the system is very low (e.g. a value of 10 or less, see Figure 2 – right). This behavior is a desired outcome based on the recommended values found in the literature[80] to apply a more systematic correction given that the PDH term tends to significantly overestimate the contribution to $\gamma_s$ and significantly underestimate the contribution to $\gamma_\pm$ when $\varepsilon_m$ or $\varepsilon_s$ are too low.[48] In this regard, details regarding the behavior of $b_{(0)}$ and $b_{(1)}$ are included in the Supporting Information. The Supporting Information also includes the corresponding plots that show a numerical evaluation for the Gibbs-Duhem consistency test with equation (10) for this IL system. Both equations (15) and (16) as well as the extension from Chang and Lin[8] were found to be Gibbs-Duhem consistent, as expected.

Regarding mixture properties, the literature states that the relative permittivity has a dominant effect for accurate calculations at high concentrations in salt and IL systems. This has been either discussed, demonstrated or directly applied to COSMO-RS based models,[8,12] UNIQUAC based models,[81–83] or the electrolyte NRTL,[14] as well as electrolyte EoS,[16,18] particularly for the electrolyte CPA equation of state by Maribo-Mogensen et al.[15,43,59,84] This effect has also been evaluated from the isolated primitive model perspective with Debye-Hückel theory by Shilov and Lyashchenko[38–40] and Lei et al.,[42] with the Monte-Carlo based IM + II model from Valiskó and Boda[44] and in the context of the Mean Spherical Approximation (MSA).[85,86] A brief analysis performed by Chang and Lin[8] arrives to the same conclusion: the behavior of the concentration dependent Debye-Hückel parameter $A_x$ and the closest approach parameter in E-PDH are dominated by their dependency on $1/\varepsilon_m$.

As a supplement to this discussion, one may isolate the contributions to the partial molar excess Gibbs free energy arising from the consistent inclusion of molar mass, the density and the relative permittivity to $ln(\gamma_\pm)$ in equation (16) with volume fraction mixing rules (equal to equation (19)) as follows:

$$\ln(\gamma_{1,i}) = -A_x \left( \frac{-2I_x^{3/2}}{(1+b''\sqrt{I_x})} \left(\frac{M_i}{M_m} - 1\right) \right) \tag{32}$$

$$\ln(\gamma_{2,i}) = -A_x \left(\frac{V_i}{V_m}\right)\left(1 - \frac{d_i}{d_m}\right)\left\{ -\frac{2I_x^{3/2}}{(1+b''\sqrt{I_x})} \right\} \tag{33}$$

$$\ln(\gamma_{3,i}) = -A_x \left(\frac{V_i}{V_m}\right)\left(\frac{\varepsilon_i}{\varepsilon_m} - 1\right) \left\{ -\frac{2I_x^{3/2}}{(1+b''\sqrt{I_x})}\left(1 + \frac{3b_{(1)}}{b''}\right) \right.$$
$$\left. -\frac{2I_x \ln\left(1+b''I_x^{1/2}\right)}{b''}\left(2 - \frac{3b_{(1)}}{b''}\right) \right\} \tag{34}$$

$$\gamma_{N,j} = \left[(\gamma_{N,cat})^{v_{cat}}(\gamma_{N,an})^{v_{an}}\right]^{\frac{1}{v_{cat}+v_{an}}} \tag{35}$$

where $v_{cat}$ and $v_{an}$ are stoichiometric coefficients. Figure 3 shows the corresponding logarithmic mean ionic averages that contribute to $\gamma_\pm$ and $\gamma_s$ for each property as functions of concentration.

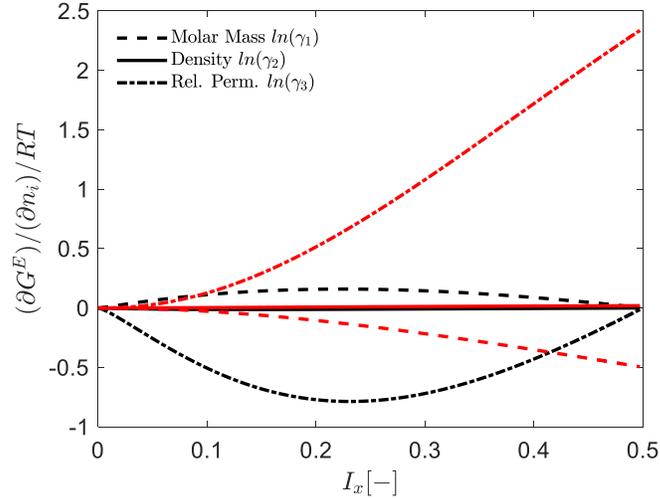

*Figure 3. Contributions from the PDH term to $\ln(\gamma_i^{PDH})$ from the concentration dependent molar mass, density and relative permittivity for aqueous [BMIM][BF₄] at room temperature calculated with a hard-core collision diameter $a = 4$ Å. Black lines are contributions to the ions, red lines are contributions to the solvent. Calculations for the $\varepsilon_{IL} = 13.9$ case.*

As expected, Figure 3 shows that the relative permittivity of the mixture has a dominant effect on the changes in qualitative behavior of the partial molar excess Gibbs free energy. Qualitative deviations are strongest for the PDH contribution to the solvent and the magnitude of these deviations directly correlate to the magnitude of the difference between $\varepsilon_s$ and $\varepsilon_{IL}$. The smaller the difference between the $\varepsilon_s$ and $\varepsilon_{IL}$, the more relevant the qualitative contribution arising from variable molar mass and density. This would be uncommon, given that ILs have an *effective* relative permittivity in a typical range[69] of $9 \leq \varepsilon_{IL} \leq 16$ which lies below that of many typical solvents. Furthermore, $\varepsilon_{IL}$ values are considered controversial in the literature,[69,87–89] given the underlying assumptions and necessary extrapolations as well as technical difficulty of measurement.[89]

Regarding the ions, by observing the contributions to $\ln(\gamma_\pm)$ from the modified extended theory, it quickly becomes evident that these contributions are rather small but not irrelevant. Qualitative deviations in practical calculations between the extended and conventional PDH terms from equations (16) and (8) can be relatively small at low to moderate concentrations if one applies concentration dependent properties in both equations. However, the trouble of adding more terms to the PDH equation and having to differentiate functions for $d_m$ and $\varepsilon_m$ is the price to pay for retaining thermodynamic consistency. Furthermore, as one approaches a highly concentrated salt regime, the cumulative difference between the values of $\frac{\partial G^E}{RT \partial n_i}$ from the extended and conventional PDH can be of one or more logarithmic units.

Finally, assuming one wishes to introduce a dissociation degree for the IL with explicit neutrally bonded ion pairs as species, then equations (17) to (19) remain unchanged. The only difference is the inclusion of additional neutrally charged molecular IL which requires a renormalization based on an educated guess[8,9,90] for the dissociation degree. Another alternative is solving the law of mass action with a known association constant and a model for the activity coefficient of the neutrally charged molecular IL (hereafter referred to as the pseudo-solvent).

Consistency with this strategy would strictly require that equation (19) yields two independent contributions for the neutral species: $ln(\gamma_s^{PDH})$ for the conventional solvent and $ln(\gamma_{s,IL}^{PDH})$ for the pseudo-solvent. Therefore, a Gibbs-Duhem consistency test applies with a form of equation (10) given by:

$$\sum_j x_j \frac{\Delta \ln(\gamma_j^{PDH})}{\Delta x_s} = -x_s \frac{\Delta \ln(\gamma_s^{PDH})}{\Delta x_s} - x_{s,IL} \frac{\Delta \ln(\gamma_{s,IL}^{PDH})}{\Delta x_s} \tag{36}$$

As a rough estimate for testing purposes, let us assume a degree of dissociation ($\xi$) of 70% for the previously described aqueous [BMIM][BF$_4$] system and that $\xi$ remains constant throughout the whole concentration range. The mixing rules remain the same and only apply a renormalized ion concentration and add an additional contribution from the corresponding degree $(1 - \xi)$ of associated pseudo-solvent. The defitions and assumptions for the calculated properties are: $M_{IL} = M_{(+)} + M_{(-)}$ for the molar mass of the pseudo-solvent. For both the pseudo-solvent and free ions we assume $d_{(+)} = d_{(-)} = d_{IL}$ for density and $\varepsilon_{(+)} = \varepsilon_{(-)} = \varepsilon_{IL}$ for relative permittivity, but $M_j = \frac{M_{IL}}{v}$ for the free ions. On a molality basis, the solvent + pseudo-solvent mass is renormalized for the 1:1 electrolyte with equation (37) and the mole fraction of the ions is renormalized with equation (38):

$$m_{s+IL} [kg] = 1 + (1 - \xi)(c_{salt}^{(m)})(M_{IL}) \tag{37}$$

$$x_j = \frac{v_j c_{salt}^{(m)} \xi}{(v_{(+)} + v_{(-)}) c_{salt}^{(m)} \xi + \frac{m_{s+IL}}{\overline{M}_{s+IL}}} \tag{38}$$

where $c_{salt}^{(m)}$ stands for total IL concentration on a molality basis, $m_{s+IL}$ stands for the combined mass of solvent and pseudo-solvent, and $\overline{M}_{s+IL}$ corresponds to the ion-free, averaged molar mass of the solvent + pseudo-solvent. The corresponding results are compared with the fully dissociated version in Figure 4 and the corresponding Gibbs-Duhem consistency test with equation (36) for the calculations with partial dissociation and neutral pseudo-solvent is included in the Supporting Information.

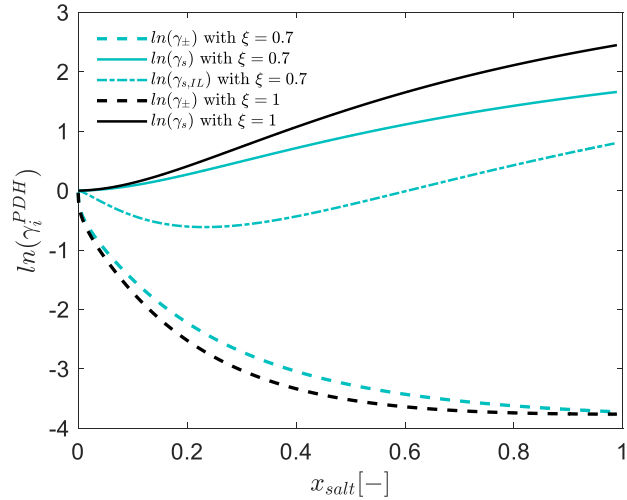

*Figure 4. Contributions to $ln(\gamma_\pm^{PDH})$ and $ln(\gamma_s^{PDH})$ for aqueous [BMIM][BF$_4$] at room temperature applying equation (19) with full ($\xi = 1$) and partial dissociation ($\xi = 0.7$) using a hard-core collision diameter of $a = 4$ Å. Handling $ln(\gamma_\pm^{PDH})$ in the conventional $ln(\xi \gamma_\pm^{PDH})$ form for explicit ion pairing also retains Gibbs-Duhem consistency.*

Figure 4 shows the qualitative impact in the activity coefficients when comparing calculations with full and partial dissociation treating a fraction of the IL as a pseudo-solvent. Several average values for the degree of dissociation are reported in the literature and have been used in thermodynamic modelling with concentration dependent properties.[8,9,14] Further systematic evaluation of different modelling strategies is required in order to assess the suitability of full and partial dissociation for ionic liquid systems whithin this framework. However, such modelling approaches are supported by the work of Chang and Lin.[8,9] In their work, the COSMO-SAC model is tested in a wide diversity of ionic liquids to calculate phase equilibria, infinite dilution activity coefficients and osmotic coefficients assuming 35% of the ionic liquid is associated and calculating the association constant based on the pure ionic liquid state.

However, DH theory considers ions that are equal in every way except charge and a dielectric medium with no other attributes than density, molar mass and relative permittivity. Therefore, while it seems tempting to take any property $\theta_i$ for a given component $i$ in equations (15) or (16), strict adherence to the DH theoretical framework requires handling averaged properties (e.g. $M_j = \frac{M_{IL}}{v}$) in the case of the ions. Splitting $\gamma_s$ into more than one contribution (due to molecular IL or to mixed solvents) is not a restriction, but requires further assessment in spite of it being a Gibbs-Duhem consistent calculation. At present, the authors suggest calculating properties for two averaged states: solvent-free and salt- (or IL-) free.

To develope a generally applicable model for ILs and regular salts over the complete comcentration range in mixed solvent systems is the goal of our work. Evaluations with extended PDH terms and extended electrolyte theories for ionic liquids are becoming common;[8,9,33] however, to the best of our knowledge, an extended, consistent PDH term has not been also systematically applied with common salt systems (e.g. alkali halides) to calculate mean ionic activity coefficients and predict phase equilibria. This is the direction and central motivation of the second and third sub-sections of the present work.

**5.2. Theoretical Primitive Model comparison for aqueous NaCl: Results**

Figure 5 shows several primitive model calculations for the more conventional aqueous NaCl system. All calculations are performed with the same Born term applied by Valiskó and Boda in their ion-ion (II) + ion-water (IW) model.[44] The II + IW model consists of a Monte-Carlo based primitive model calculation for the II term and a Born term for transfer between infinite dilution and a concentrated solution for the IW term. The unrestricted hard-sphere term is in good agreement with Monte-Carlo values[48,74] and the effect of size asymmetry between the ions is dominated by this term at low to moderate concentrations.[48] It is therefore safe to assume that the electrostatic long-range contribution of the PDH term is majorly responsible for any qualitative differences with respect to the II + IW model and experimental data.

Based on the aforementioned assumption, one can observe in Figure 5 that equation (15), the E-PDH term with $b'$, requires an empirical adjustment to fit the values of $\ln(\gamma_\pm)$ with reference values. If $b'$ is left unchanged it incurs into a systematic underestimation of theoretical origin: the Debye-Hückel derivation[21] via the Guntelberg charging process is not in agreement with its equivalent derivation[24] via the pressure equation; a well-known inconsistency[62] that is addressed from the statistical mechanics stand-point by the recently published Multiple Decay-Length Extension of the Debye-Hückel theory.[56] The historical workaround for practical application in chemical engineering is either to fit the parameter of closest approach to experimental data or to scale the hard-core collision diameter by a constant (if not left as an open parameter). These options are effective because the parallel capacitor configuration of all linearized theories allows them to behave similarly when scaling the hard-core collision diameter.[48]

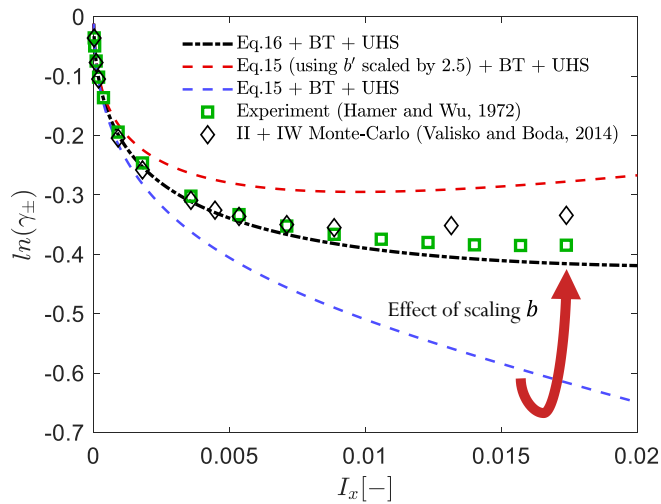

*Figure 5. Left: calculations for $ln(\gamma_\pm)$ [mole fraction scale] of aqueous NaCl (T=25°C) evaluated with equation (15) and with equation (16). The electrostatic PDH terms are combined with an unrestricted hard-core contribution (UHS) and the Born term (BT) from the IS+II Model as previously described. Experimental data from Hamer and Wu[73] is shown for reference.*

One may also observe in Figure 5 that scaling $b'$ by 2.5 incurs into a systematic overestimation with respect to the reference values. In order for equation (15) to be in good agreement with the reference values, one would have to scale $a$ by a value of 1.8, which corresponds to a parameter of closest approach parameter of roughly $b'^{(m)} = 1.5$ in the molality convention and lies closer to literature values[24,65] for water. In contrast, it can be observed in Figure 5 that equation (16) with $b''$ provides a good qualitative fit with the additional benefit that it will provide a systematic adjustment as the relative permittivity of the mixture changes when compared to equation (15) using $b'$. The PDH term requires this systematic adjustment, as demonstrated historically by its literature values. It is therefore recommended to use equation (16) or alternatively to use equation (15) with $b'$ scaled by a constant of 1.5 or larger depending on modelling requirements. However, for systems with a wide range of relative permittivity values, including (as rule of thumb) $\varepsilon_r < 10$ for the mixture, or systems with a pronounced dielectric decrement, the use of the modified extension from equation (16) should be favored.

As shown in our previous work,[48] purely mathematical calculations in the primitive model that do not account for ion pairing systematically overestimate the values of $\gamma_\pm$. This has been a well-known fact for decades[91] and it must be clarified that equation (13) applies a systematic correction for a varying and/or very low relative permittivity and therefore correlates to some extent with ion pairing, but it does not account for such a phenomenon explicitly. The result shown in Figure 5 is semi-empirical since $b''$ seeks to reproduce literature values. Therefore, the modified parameter of closest approach can be combined and balanced with a model that accounts for ion-ion and ion-solvent short-ranged interactions, like NRTL, UNIQUAC or UNIFAC based models, or in this case the COSMO-RS-ES model,[10–12] provided that concentration dependent properties are reasonably described.

### 5.3. Universal application for common salts with the COSMO-RS-ES model: Results

Having validated equations (15) and (16), these can be combined with the COSMO-RS-ES model as new long-range terms that replace the traditional PDH term from equation (8). Table 3 shows the general results of the average absolute deviation (AAD) obtained from several parameterizations of the model when calculating aqueous, mixed and non-aqueous MIACs as well as LLE data and dielectric decrement (when applicable). Salt solubility predictions in mixed aqueous and non-aqueous solvents are also included. The long-range PDH term applied in each parameterization and its corresponding parameter of closest approach are shown in the upper part of the table. It is also indicated whether the data belongs or not to the training set of each parameterization. One must stress the fact that in this case, solubility predictions do not require any system or component specific binary interaction parameters and that the form of the short-ranged COSMO-RS based term has not been modified. Thus, any improvements in overall correlative and predictive capabilities are attributed to the modification and/or extension of the long-range PDH term.

One may conclude from Table 3 that parameterization C, which applies ME-PDH from equation (16) as long-range term, outperforms parameterization B, which applies the E-PDH extension from equation (15). This is an expected result given that the calculation assumes full dissociation and several experimental data points correspond to pure or mixed solvent phases with a very low relative permittivity (e.g. 2-methyl-2-butanol $\varepsilon_s = 5.8$, dimethyl ether $\varepsilon_s = 5$, dimethyl carbonate $\varepsilon_s = 3.1$). Thus, in agreement with the recommended literature values for the parameter of closest approach, the ME-PDH term in the present calculations provides a correction given by $b''$, which results in more adequate estimates of $\gamma_\pm$. This systematic correction applies not only for low relative permittivity values of the solvent, as shown our previous work,[48] but also for a better qualitative performance when introducing the dielectric decrement, as previously discussed and shown in Figure 5 for primitive model calculations and supported by the results of the present section.

The obtained factor $f_{sc}$, which scales the sum of the COSMO radii applied in the PDH term to calculate $a$, is larger for parameterization B than for parameterization C (1.09 vs 0.63). This means that the optimization algorithm improved modelling results in parameterization B by scaling the value of $b'$, thereby suggesting that the use of E-PDH requires (on average) larger values for the collision diameter $a$ when compared to ME-PDH. This partially explains the empirical pre-factor of $2.5 \times a$ in the work of Chang and Lin[8] for the calculation of $b'$. Nevertheless, the present results suggest a better overall performance when favoring the use of $b''$ over $b'$ in the PDH extension for the correlation of a large, varied database.

*Table 3. Results of the COSMO-RS-ES model parameterized with equations (8), (15) and (16) as long-range electrostatic contributions. Equation (31) is used for the effect of ions as point charges on the relative permittivity of the medium.*

| | Parameterization | A [a] | B | C | D [b] |
|---|---|---|---|---|---|
| | PDH term | PDH Equation (8) (salt-free properties) | E-PDH Equation (15) (conc. dep. properties) | ME-PDH Equation (16) (conc. dep. properties) | ME-PDH Equation (16) (conc. dep. properties) |
| | Optimized parameters in the description of $\varepsilon_m$ | N/A | $\alpha_{salt} = 1.82$ | $\alpha_{salt} = 2.39$ | $\alpha_{salt} = 2.38$ $\varepsilon_{salt} = 30.8$ |
| | Parameter of closest approach | 14.9 | $b'$ | $b''$ | $b''$ |
| | System Type and number of experimental data points | Average Absolute Deviation(AAD) | | | |
| Part of the Training Set | MIAC (aqueous + mixed aqueous solvents, $N_{DP}$ = 2889) | 0.090 | 0.111 | 0.076 | 0.069 |
| | MIAC (non-aqueous, $N_{DP}$ = 976) | 1.950 | 0.835 | 0.434 | 0.497 |
| | LLE ($N_{DP}$ = 1056) | 0.794 | 0.781 | 0.718 | 0.660 |
| | Dielectric decrement ($N_{DP}$ = 259) | N/A | 6.281 | 5.201 | N/A |
| **Not** part of the Training Set | Dielectric decrement ($N_{DP}$ = 259) | N/A | N/A | N/A | 5.918 |
| | SLE ($N_{DP}$ = 836) | 0.923 | 0.905 | 1.027 | 0.948 |
| **TOTAL** [c] | $N_{DP}$ = 5757 [c] | 0.655 | 0.473 | 0.392 | 0.375 |

[a] Assumes salt free properties ($\varepsilon_s, d_s, M_s$) with the conventional PDH term from equation (8) and $b = 14.9$. [b] Applies $\varepsilon_{salt}$ as an additional parameter instead of a constant $\varepsilon_{salt} = 13$ and excludes the dielectric decrement as part of the training set. [c] Total AAD of MIAC, LLE and SLE systems, excluding the AAD of the dielectric decrement calculations. N/A = Not Applicable.

As a general finding of the present work, it is shown that by means of an extended electrostatic theory, the COSMO-RS-ES model can simultaneously correlate low and high salt concentrations in aqueous systems with an accuracy that is comparable to its preceding[10–12] versions (which were limited to a molality of 6). However, performance for non-aqueous systems is considerably superior. As evidenced in the results from Table 3, the capabilities of the model to correlate this type of data in combination with aqueous data improve considerably, but still have room for improvement: calculations for aqueous systems remain much more accurate when compared to non-aqueous systems.

So far, the model applies specific parameters for water to ion related interactions, whereas universal parameters for all organic solvents are used, making no distinctions between strong, weak or more polarizable donors and acceptors (see Table 2). It is therefore an expected result that the model favors aqueous systems over non-aqueous systems. Such an observation points towards future opportunities for model development, particularly for COSMO-RS based ion to solvent interactions that make distinctions between multiple segment descriptors. Nevertheless, the present work does not focus on COSMO-RS-ES, but on an isolated, generalized improvement of long-range electrostatics alone, that can be applicable to any other $g^E$-model or EoS for electrolytes. As such, the results from Table 3 are to be interpreted in terms of a predictive $g^E$-model model being used as a tool to discuss the benefits of the extended and modified extended PDH terms with respect to the conventional PDH term. Improvements or modifications of the short-range COSMO-RS based term remain object of future work.

Moving on to the use of concentration dependent properties, calculations performed with hard-core and Born terms in our previous work,[48] as well as in section 5.2 of the present work, apply raw experimental data for the dielectric decrement. This has also been the selected modelling strategy in theoretical works like those of Shilov and Lyashchenko[38–40] or Valiskó and Boda.[44] Nevertheless, the experimentally measured dielectric decrement has a kinetic contribution.[92,93] Whether its effect is small or not remains a discussed topic in the literature.[94] It is sometimes not considered when Born radii are being fitted along with a hard-sphere and electrostatic terms.[40] The case could be different for a chemical engineering model. For instance, in the work of Maribo-Mogensen et al.[15,84] with the electrolyte CPA equation of state, the effect of kinetic depolarization in a chemical engineering thermodynamic model is discussed. As performed in their work, and suggested in the theories of Hubbard and Onsager,[92,93] one may estimate the kinetic contribution with the no-slip condition using equation (39):

$$\Delta\varepsilon_k = \left(1 - \frac{\varepsilon_\infty}{\varepsilon_{exp}}\right)\frac{\tau \Lambda_{el}}{\varepsilon_0} \tag{39}$$

where the quantity $\Delta\varepsilon_k$ must be removed from the experimentally measured decrease of the relative permittivity of the mixture $\varepsilon_{exp}$. The quantity $\varepsilon_\infty$ is the zero frequency permittivity of the solvent, $\tau$ is the relaxation time of the solvent, $\Lambda_{el}$ is the conductivity of the electrolyte and $\varepsilon_0$ is vacuum permittivity. For the perfect slip condition, equation (39) is scaled by a factor of $2/3$.[93]

The kinetic contribution must be discussed given that for parameterizations B and C there is an average absolute deviation of 6.281 and 5.201 for the dielectric decrement, respectively, and these deviations arise from systematic overestimation of the fitted equation (31) with respect to experimental values. This overestimation of the dielectric decrement can be (at least partially) attributed to kinetic depolarization. Consequently, an additional parameterization (parameterization D) was performed. Parameterization D is in essence a refinement of Parameterization C in the sense that it excludes the dielectric decrement data as a constraint in the training set (making $\varepsilon_m$ an output instead of a correlation) and allows the hypothetical limiting value $\varepsilon_{salt}$ to be an optimizable parameter, as stated in Table 3. The parameters for parameterization D and parameterizations B and C are included in the Supporting Information.

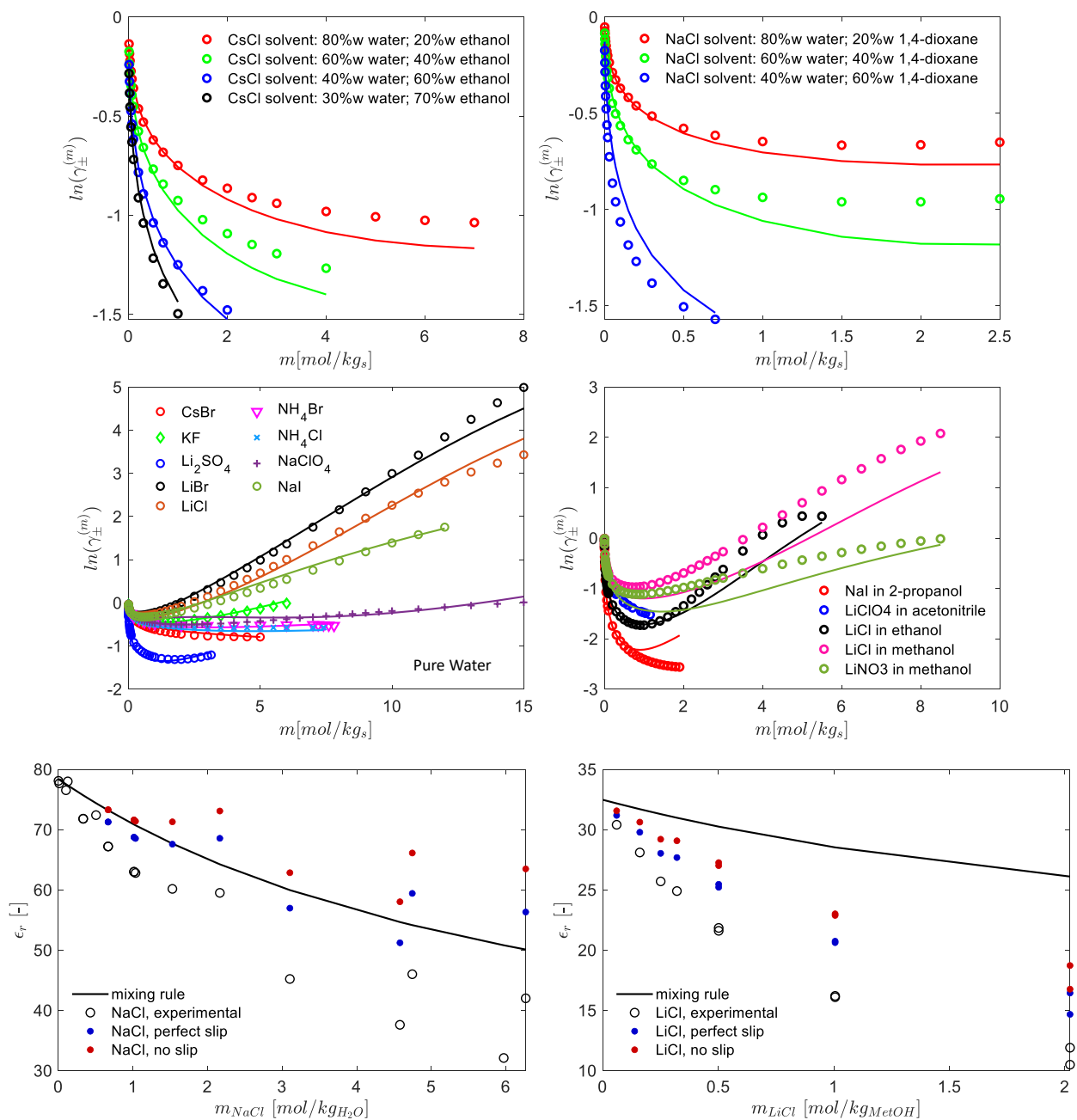

*Figure 6. Calculations with Parameterization D. Top and center: MIAC calculations for CsCl in a water/ethanol mixture, sodium salts in pure organic solvents, NaCl in a water/1,4-dioxane mixture and moderate to highly soluble salts in pure water and pure alcohols. Bottom plots: relative permittivity of NaCl in water and LiCl in methanol. Open symbols are either experimental data or a Pitzer-Myorga fit of the experimental osmotic coefficient data. Sources: Bonner et al.[95], Barthel et al.[96], Nasirzadeh & Neueder,[97] Safarov et al.[98], Ceccattini et al.[99], Hamer and Wu[73], Goldberg[100], Winsor and Cole,[101] Kaatze et al.,[102] Hasted et al.[103], Hasted and Roderick,[104] and Harris and O'Konski.[105] For the perfect slip and no-slip conditions, the calculations of $\Delta\varepsilon_k$ applied $\tau = 7.4$ [ps] for water and $\tau = 50$ [ps] for methanol. Conductivity data for the $\Delta\varepsilon_k$ calculations was correlated from Chen et al.,[106] Wu et al.,[107] and Bešter-Rogač et al.[108]*

MIAC calculations and dielectric decrement correlations from parameterization D are shown in Figure 6 for several systems. In Figure 6 (top and center), one may observe good qualitative and quantitative agreement for aqueous systems including those with a strongly variable salt-free relative permittivity like the NaCl/water/1,4-dioxane (1,4-dioxane has a relative permittivity close to 2.3). As previously discussed, these systems are being successfully described for the first time with concentration dependent relative permittivity, density and molar mass of the mixture within the model.

Qualitative trends also hold for non-aqueous systems, but some larger systematic deviations can be observed. Additional considerations are required in order to improve the modelling of non-aqueous systems even further and these can go beyond the addition of parameters in the short-range term i.e. the presence of extensive ion pairing, which is not considered in the present work. For instance, calculations for NaI + 2-propanol in Figure 6 – center (right) lack the typical effect of ion pairing that pulls $\ln(\gamma_\pm)$ values down. Nevertheless, the use of ME-PDH by itself already provides a significant improvement.

The dielectric decrement plots include an estimation for kinetic depolarization effects estimated with equation (39) applying both the no-slip and perfect slip conditions. In the case of aqueous systems where enough data was available to evaluate equation (39), kinetic effects, as exemplified for aqueous NaCl in Figure 6 – bottom (left), could explain the systematic overestimation of the concentration dependent relative permittivity (two additional examples are included in the Supporting Information). Nevertheless, Figure 6 – bottom (right) shows that this is not entirely the case for non-aqueous systems, as shown for LiCl in methanol. In this regard, the mixing rule from equation (31) applied in parameterization D predicts a value of $\varepsilon_{salt} = 30.8$. This seems to agree with measured values of the permittivity of some ionic liquids,[69,89] averaged estimates[8] for $\varepsilon_{IL}$ and extrapolated effective values in the modelling of aqueous electrolytes with conventional salts[109] that reach values of 20 or higher for $\varepsilon_{IL}$ (equivalent to $\varepsilon_{salt}$ as hypothetical extrapolations). Nevertheless, these experimental and hypothetical values serve more as useful tool for modelling purposes notwithstanding that additional information on the fluid structure throughout the whole concentration range is required and that limiting values of $\varepsilon_m$ are strongly solvent and salt specific,[109,110] which explains the stronger deviation for the dielectric decrement in methanol. Limiting values for solvent specific cases[84,109] could provide improvement in future model development.

Overall, kinetic depolarization constitutes a good plausible explanation of why $\varepsilon_m$ is overestimated with respect to experimental values and why the predicted value of $\varepsilon_{salt}$ is higher than the reported values. Structural changes from ion-ion and ion-solvent interactions, the effect of ion pairing and changes in the solvation shell are still needed to be understood better for a more complete description of $\varepsilon_m$ and $\frac{\partial \varepsilon_m}{\partial n_i}$. Thus, we emphasize the need for additional efforts to understand and model the relative permittivity of electrolyte mixtures.

Furthermore, dielectric decrement data for both salt and ionic liquid systems is scarce, highly variable and controversial[69,89] and the modelling of the relative permittivity of highly polar solvents without ions is already a challenging topic.[111] Interestingly, modified mixing rules like equation (31) and others applied in the literature[8,16,18,68] seem to offer a rough, but acceptable general approach to this problem as long as precision in the modelling of $\varepsilon_m$ is not a priority and additional parameters in a model are able to counter-balance systematic deviations. One must stress the fact that combining long-range electrostatics as previously described with a predictive model like COSMO-RS-ES will challenge the capabilities of the short-range model to describe qualitative trends. Consequently, further model development is of paramount relevance in order to achieve a better coupling between COSMO-RS based calculations and the behavior of $\varepsilon_m$, particularly in systems with small, strongly solvated ions.

It is also relevant to state that COSMO-RS-ES, as any other conventional g$^E$-model, has no access to the density or structure of the fluid and relies on experimental data, QSPR methods and mixing rules to describe $\varepsilon_m$ and $d_m$. At present, terms like E-PDH or ME-PDH could offer a self-consistent approach if coupled with an electrolyte EoS like the e-CPA version from the DTU group,[15,84] which describes system specific $d_m$ and works with a statistic-mechanical method to describe $\varepsilon_m$ (and their partial derivatives) in terms of pressure, temperature and concentration. Such a strategy could potentially provide further insights into the use of concentration dependent properties electrolyte modelling.

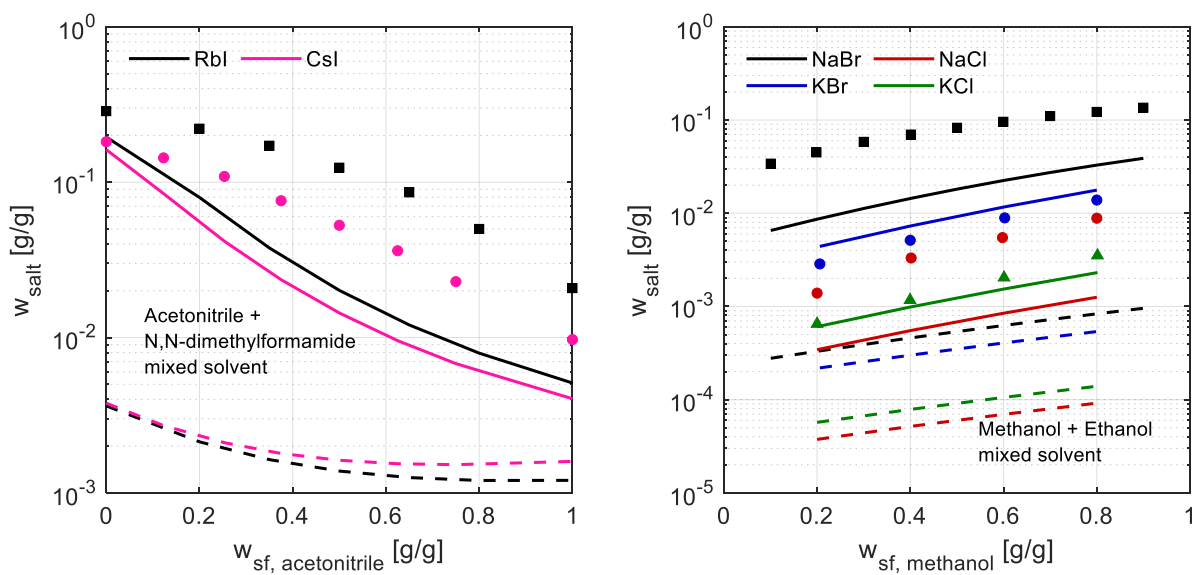

*Figure 7. Predictions with parameterization D (full-lines) compared to a previous COSMO-RS-ES version from Müller et al.[12] (dashed lines). Left: solubility of CsI and RbI in an acetonitrile/N,N-dimethylformamide mixed solvent. Experimental data from the IUPAC-NIST Solubility Data Series.[112] Right: Predictions of the solubility of several salts in a non-aqueous methanol / ethanol mixture. Experimental data from Pinho and Macedo.[32,113] Symbols are experimental data.*

Moving on to predictive power, the reliable capabilities of the COSMO-RS-ES model for aqueous and mixed aqueous electrolytes was previously discussed and has been demonstrated in other works.[10–12] However, highly concentrated, non-aqueous electrolytes remain a considerable challenge for electrolyte models in general, particularly when handling salt solubility.[114] In this work, advances are achieved within the COSMO-RS-ES framework. Figure 7, for instance, presents full predictions of the solubility of RbI and CsI in an acetonitrile/N,N-dimethylformamide mixed solvent and of four other salts in an ethanol/methanol mixed solvent. One may observe that the predictions from parameterization D (full lines) outperform predictions from a previous refined version[12] of the COSMO-RS-ES model (dashed lines) which used the traditional PDH term with a renormalized effective ionic strength.

While both parameterizations of the model have a measure of qualitative agreement with experimental trends, the preceding version underestimated the solubility by one to two orders of magnitude in all the systems shown in Figure 7 and had poorer qualitative agreement in Figure 7 – left. In contrast, the present work provides predictions that are quantitatively and qualitatively more reliable. While COSMO-RS-ES evidently has opportunity for improvement for non-aqueous systems, evidence of the systematic improvement through additional considerations in long-range electrostatics alone is provided in a fashion that could be extrapolated to any model.

Revisions of the COSMO-RS based short-range contribution are a separate topic and scope of future work. The improvements demonstrated in Figure 7 are therefore attributed to the use of the ME-PDH term from equation (16). Limiting the description of $\varepsilon_m$ and applying the same short-range contribution from previous works[12,48] strengthens the argument that applying a modified and extended long-range PDH term with concentration dependent properties translates into improvements for practical application. Extension and modification of the electrostatic theory alone leads to a better performance when compared to just enlarging the dataset (as demonstrated in the comparison from Table 3) or modifying the modelling approach with an incomplete physical description (as demonstrated in the comparison from Figure 7). In this regard, there is a current need for consistent, standardized thermodynamic models that effectively handle non-aqueous electrolyte systems more effectively.[1,2,114] The present work suggests that this challenge can be approached not only with generalized extensions, but sometimes also with semi-empirical modifications of the commonly applied electrolyte theories provided that thermodynamic consistency is retained.

*Table 4. Summary of modified, Gibbs-Duhem consistent PDH terms (unsymmetric). The modified parameter of closest approach $b''$ is given by equation (11).*

| **Modified PDH + Extensions validated for single salt (or single IL) systems. Unsymmetric reference state.** | **Comments**[a] |
|---|---|
| $$\ln(\gamma_i^{M-PDH}) = -A_x \left\{ \frac{2z_i^2}{b''} \ln(1 + b''\sqrt{I_x}) + \frac{z_i^2 I_x^{1/2} - 2I_x^{3/2}}{(1 + b''\sqrt{I_x})} \right\}$$ | Modified PDH (M-PDH) with no extension. Takes salt-free properties $M_s, d_s, \varepsilon_s$ when calculating $A_x$ and $b''$. Can replace the PDH term in published $g^E$-models. Recommended for dilute and moderately concentrated systems. An equivalent version was validated in our previous work.[48] |
| $$\ln(\gamma_i^{ME-PDH}) = -A_x \left\{ \frac{2z_i^2}{b''} \ln(1 + b''\sqrt{I_x}) + \frac{z_i^2 I_x^{1/2} - 2I_x^{3/2}\left(\frac{M_i}{M_m}\right)}{(1 + b''\sqrt{I_x})} - \frac{2I_x^{3/2} n_T}{(1 + b''\sqrt{I_x})} \left[ \frac{1}{\varepsilon_m}\left(1 + \frac{3b_{(1)}}{b''}\right)\frac{\partial \varepsilon_m}{\partial n_i} - \frac{1}{d_m}\frac{\partial d_m}{\partial n_i} \right] \right.$$ $$\left. - \frac{2I_x \ln(1 + b'' I_x^{1/2}) n_T}{b'' \varepsilon_m}\left(2 - \frac{3b_{(1)}}{b''}\right)\frac{\partial \varepsilon_m}{\partial n_i} \right\}$$ | Fully generalized ME-PDH: takes variable $M_m, d_m, \varepsilon_m$ when calculating $A_x$ and $b''$. Applicable to any system if experimental data or concentration dependent definitions for $\varepsilon_m$ and $d_m$ are available. |
| $$\ln(\gamma_i^{ME-PDH})_\phi = -A_x \left\{ \frac{2z_i^2}{b''} \ln(1 + b''\sqrt{I_x}) + \frac{z_i^2 I_x^{1/2} - 2I_x^{3/2}\left(\frac{M_i}{M_m}\right)}{(1 + b''\sqrt{I_x})} - \frac{2I_x^{3/2}}{(1 + b''\sqrt{I_x})}\left[\left(\frac{V_i}{V_m}\right)\left(1 - \frac{d_i}{d_m}\right) + \left(\frac{V_i}{V_m}\right)\left(1 + \frac{3b_{(1)}}{b''}\right)\left(\frac{\varepsilon_i}{\varepsilon_m} - 1\right)\right] \right.$$ $$\left. - \frac{2I_x \ln(1 + b'' I_x^{1/2})}{b''}\left(2 - \frac{3b_{(1)}}{b''}\right)\left(\frac{V_i}{V_m}\right)\left(\frac{\varepsilon_i}{\varepsilon_m} - 1\right) \right\}$$ | ME-PDH using volume fraction mixing rules (denoted by subscript $\phi$) for $\varepsilon_m$ and $d_m$. Takes variable $M_m, d_m, \varepsilon_m$ when calculating $A_x$ and $b''$. A somewhat similar E-PDH based term has been successfully tested in the literature by Chang and Lin[8,9] and Ganguly et al.[33] for ionic liquid systems. Applicable to polyatomic ions with delocalized charges. |

[a] *The recommended values for equations (11) to (13) are $\omega_{(0)} = 3/2$ and $\omega_{(1)} = 1/9$ with an averaged hard core collision diameter $a \approx 3$ Å based on the radii from Marcus.[70] In all cases, applying $b''$ with $\omega_{(0)} = 1$ and $\omega_{(1)} = 0$ results in the homologue relations for the PDH or E-PDH terms for which $a \approx 4.25$ Å is the recommended average value[48] for conventional salts. Volume and mass fraction mixing rules are directly interchangeable in the ME-PDH $\phi$ extension by the use of molar volumes $V_i/V_m$ or molar masses $M_i/M_m$ as pre-factors, respectively. The volume fraction approach assumes that the molar volumes of the ions in the solution remain constant. The use of averaged properties for all ions is suggested: $\varepsilon_j = \varepsilon_{IL}$, $d_j = d_{IL}$, $V_j = \frac{V_{IL}}{\nu}$ and $M_j = \frac{M_{IL}}{\nu}$ where subscript j stands for an arbitrary ion. Effective values or educated guesses are required for $\varepsilon_{IL}$ and $d_{IL}$ (or hypothetical limiting values for conventional salts with a solubility limit).*

Table 5. Summary of modified, Gibbs-Duhem consistent PDH terms (symmetric). The modified parameter of closest approach $b''$ is given by equation (11).

| Modified PDH + Extensions validated for single salt (or single IL) systems. Symmetric reference state. | Comments[a] |
|---|---|
| $$\ln(\gamma_{i,\ symm}^{M-PDH}) = -A_x \left\{ \frac{2z_i^2}{b''} \ln\left(\frac{1+b''\sqrt{I_x}}{1+b''\sqrt{I_x^{IL}}}\right) + \frac{z_i^2 I_x^{1/2} - 2I_x^{3/2}}{(1+b''\sqrt{I_x})} - \frac{z_i^2 \sqrt{I_x^{IL}} - 2(I_x^{IL})^{3/2}}{1+b''\sqrt{I_x^{IL}}} \right\}$$ | $A_x$ and $b''$ remain constant (take salt-free properties) <br><br> $I_x^{IL} = \frac{\delta_i}{2} \sum_i \frac{v_i}{v}$ $\delta_i = \begin{cases} 0, & i=s \\ 1, & i=j \end{cases}$ |
| $$\ln(\gamma_{i,\ symm}^{ME-PDH}) = -A_x \left\{ \frac{2z_i^2}{b''} \ln(1+b''\sqrt{I_x}) + \frac{z_i^2 I_x^{1/2} - 2I_x^{3/2}\left(\frac{M_i}{M_m}\right)}{(1+b''\sqrt{I_x})} - \frac{2I_x^{3/2} n_T}{(1+b''\sqrt{I_x})} \left[\frac{1}{\varepsilon_m}\left(1+\frac{3b_{(1)}}{b''}\right)\frac{\partial \varepsilon_m}{\partial n_i} - \frac{1}{d_m}\frac{\partial d_m}{\partial n_i}\right] \right.$$ $$\left. - \frac{2I_x \ln(1+b''\sqrt{I_x}) n_T}{b'' \varepsilon_m}\left(2-\frac{3b_{(1)}}{b''}\right)\frac{\partial \varepsilon_m}{\partial n_i} \right\} + A_x^{IL}\left\{\frac{2z_i^2}{b_{IL}''}\ln\left(1+b_{IL}''\sqrt{I_x^{IL}}\right) + \frac{z_i^2\sqrt{I_x^{IL}} - 2(I_x^{IL})^{3/2}}{1+b_{IL}''\sqrt{I_x^{IL}}}\right\}$$ | $A_x$ and $b''$ are salt/IL concentration dependent <br><br> $I_x^{IL} = \frac{\delta_i}{2}\sum_i \frac{v_i}{v}$ $\delta_i = \begin{cases} 0, & i=s \\ 1, & i=j \end{cases}$ <br><br> $A_x^{IL} = \frac{1}{3}\left(\frac{2\pi N_A d_{IL}}{M_{IL}}\right)^{0.5}\left(\frac{e^2}{4\pi\varepsilon_0 \varepsilon_{IL} k_B T}\right)^{1.5}$ |
| $$\ln(\gamma_{i,\ symm}^{ME-PDH})_\phi = -A_x \left\{ \frac{2z_i^2}{b''}\ln(1+b''\sqrt{I_x}) + \frac{z_i^2 I_x^{1/2} - 2I_x^{3/2}\left(\frac{M_i}{M_m}\right)}{(1+b''\sqrt{I_x})} - \frac{2I_x^{3/2}}{(1+b''\sqrt{I_x})}\left[\left(\frac{V_i}{V_m}\right)\left(1-\frac{d_i}{d_m}\right) + \left(\frac{V_i}{V_m}\right)\left(1+\frac{3b_{(1)}}{b''}\right)\left(\frac{\varepsilon_i}{\varepsilon_m}-1\right)\right] \right.$$ $$\left. - \frac{2I_x \ln(1+b''\sqrt{I_x})}{b''}\left(2-\frac{3b_{(1)}}{b''}\right)\left(\frac{V_i}{V_m}\right)\left(\frac{\varepsilon_i}{\varepsilon_m}-1\right) \right\} + A_x^{IL}\left\{\frac{2z_i^2}{b_{IL}''}\ln\left(1+b_{IL}''\sqrt{I_x^{IL}}\right) + \frac{z_i^2\sqrt{I_x^{IL}} - 2(I_x^{IL})^{3/2}}{1+b_{IL}''\sqrt{I_x^{IL}}}\right\}$$ | $b_{IL}''$ calculated using $\lambda_B^{IL}$ and $b_{IL}'$: <br><br> $\lambda_B^{IL} = \frac{e^2}{4\pi\varepsilon_0 \varepsilon_{IL} k_B T}$ <br><br> $b_{IL}' = a\sqrt{\frac{2d_{IL}N_A e^2}{M_{IL}\varepsilon_0 \varepsilon_{IL} k_B T}}$ |

[a] The recommended values for equations (11) to (13) are $\omega_{(0)} = 3/2$ and $\omega_{(1)} = 1/9$ with an averaged hard core collision diameter $a \approx 3$ Å based on the radii from Marcus.[70] In all cases, applying $b''$ with $\omega_{(0)} = 1$ and $\omega_{(1)} = 0$ results in the homologue relations for the PDH or E-PDH terms for which $a \approx 4.25$ Å is the recommended average value[48] for conventional salts. Volume and mass fraction mixing rules are directly interchangeable in the ME-PDH $\phi$ extension by the use of molar volumes $V_i/V_m$ or molar masses $M_i/M_m$ as pre-factors, respectively. The $\phi$ approach assumes that the molar volumes of the ions in the solution remain constant. Effective values or educated guesses are required for $\varepsilon_{IL}$ and $d_{IL}$ (or hypothetical limiting values for conventional salts with a solubility limit). Subscript $i$ stands for any component, subscript $j$ stands for ions, subscript $s$ stands for solvent(s). The term $v_j$ stands for the stoichiometric coefficient of an ion in the salt dissociation reaction and $v = v_{cation} + v_{anion}$. The term $v_i$ is zero when $i = s$. Relevant notes: averaged properties for the solvent-free state ($\varepsilon_j = \varepsilon_{IL}$, $d_j = d_{IL}$, $V_j = \frac{V_{IL}}{v}$ and $M_j = \frac{M_{IL}}{v}$) are mandatory for the symmetric formulation to be in agreement with the unsymmetric formulation. These equations are valid for both symmetric and unsymmetric salts. See Supporting information for details and derivation.

# 6. Conclusions

A Gibbs-Duhem consistent Pitzer-Debye-Hückel term with a systematic correction for very low and variable relative permittivity media is presented and extended for concentration dependent density, relative permittivity and molar mass. The extension is validated for electrolytes with small or large ions in a concentration range that spans from infinite dilution to the ionic liquid state and does not require any system specific adjustments for strongly variable or very low relative permittivity values.

The correction is achieved through a modified parameter of closest approach, which consists of a simplified, easily differentiable form of a preceding version and is given by the sum of two contributions with two global parameters. These terms retain thermodynamic consistency and qualitatively reproduce the trends of recommended literature values for the Pitzer equations. The first term consists of the original parameter of closest approach scaled by a universal constant to achieve agreement with MDE-DH theory when applying the same hard-core collision diameter; the second one is the original parameter of closest approach scaled by the fluid specification and its qualitative behavior is therefore dominated by the Bjerrum length as a pre-factor.

Sample calculations with the novel modified extended PDH (ME-PDH) term are performed and compared to the conventional extended PDH (E-PDH). It is shown that the conventional extension incurs in systematic deviations which the modified extension is intended to correct. Both PDH extensions are subsequently applied with concentration dependent relative permittivity, density and molar mass in combination with the COSMO-RS-ES model to correlate diluted and concentrated electrolyte systems and predict the solubility of salts in mixed aqueous and mixed non-aqueous solvents. We conclude that the use of the modified-extended PDH term results in improved overall model performance when compared with the use of the unmodified extended PDH term as well as the traditional PDH term. Particular improvement is observed in the prediction of the solubility of salts in non-aqueous media.

The averaged mixing rule applied to estimate the concentration dependent relative permittivity of the mixture in the COSMO-RS-ES based calculations yields a systematic overestimation with respect to experiment values of the dielectric decrement from the literature. Kinetic depolarization effects provide a plausible explanation for this overestimation, but further research and modelling efforts are required to describe the dielectric decrement in electrolytes in a generalized manner.

Finally, and in line with findings from the recent literature, our results support the consistent introduction of concentration dependent properties within long-range electrostatics in order to improve overall model performance when handling concentrated, non-aqueous electrolytes.

## 7. Acknowledgements

The authors are grateful to Frank Eckert, Michael Diedenhofen and the team from BIOVIA Dassault-Systèmes for the scientific discussions. AGC would like to express gratitude to Roland Kjellander for the enriching correspondence on electrolyte theory. The authors acknowledge financial support from BIOVIA Dassault-Systèmes.

## 8. Conflicts of Interest

There are no conflicts of interest to declare.